# SMIE: Weakness is Power!

## Auto-indentation with incomplete information

Stefan Monnier [a]

a   Université de Montréal, Canada

**Abstract**   Automatic indentation of source code is fundamentally a simple matter of parsing the code and then applying language- and style-specific rules about relative indentation of the various constructs. Yet, in practice, full parsing is not always an option, either because of quirks of the language, or because the code is temporarily syntactically incorrect, or because of an incomplete or broken grammar.

   I present the design of Emacs's Simple-Minded Indentation Engine (SMIE), which gets its power from the weakness of the underlying parsing technique. It makes it possible to perform *local* parsing, which is hence unaffected by irrelevant surrounding code. This provides a form of graceful degradation in the face of incomplete, erroneous, or just plain problematic information.

## The Art, Science, and Engineering of Programming

# 1 Introduction

One of the basic functionalities a programmer expects from a text editor is to help keep the code properly indented. In the context of an Integrated Development Environment (IDE), i. e. an editor with direct access to the implementation of the programming language, this can be done by consulting the syntax tree to find the relevant context of the chunk of code one needs to (re)indent, and then apply the relevant indentation rules. Of course, since we are in the process of editing the source code, we need to behave sanely even when the source code is not currently syntactically correct, which requires a well-crafted parser with good error recovery.

Access to an implementation of the language is often necessary to be able to reliably build a complete syntax tree, for example because of the language's preprocessing or metaprogramming facilities, and it can bring very significant benefits to the end user (see for example [4]). Yet there are many cases where this approach is not an option.

Text editors such as Emacs, are not tightly integrated with any particular language implementation and hence do not have easy access to a syntax tree. There, auto-indentation has traditionally worked very differently, relying on messy ad-hoc algorithms and dark magic that, by and large, noone designed nor understands. The upside is that they tend not to be affected very much by code that is syntactically incorrect, because they only search for local clues to decide how to indent a given chunk of code, instead of relying on a complete parse; their most evident downside is that they invariably only handle correctly a small subset of the language's syntax, but their main problem is the code's complexity and unmaintainability.

Emacs's Simple-Minded Indentation Engine (SMIE) brings long-overdue light to those ad-hoc algorithms, helping structure the code and make it more maintainable. It does this by taking a middle-ground approach which does parse the source code but, unlike an IDE, parses it *locally* and *backwards*, which makes it naturally error-tolerant since it does not even look at the parts of the source code that are not relevant.

# 2 Automatic indentation

Automatically indenting code can mean a few different things, but in this article we will focus on the following sub-problem: given a source code file where we assume that lines 1 to $N$ are properly indented already, compute the column to which the code on line $N + 1$ should be indented. In this section, we present the problem and the usual approaches taken to solve it, as well as the difficulties they face.

## 2.1 Indentation rules

Since indentation usually reflects the syntactic structure of the source code, automatic indentation usually involves finding the siblings or parents of the token that starts on line $N + 1$ and then applying ad-hoc rules indicating whether to align with the sibling, or indent with some offset relative to the immediate parent. Of course, in some cases, the rules can require looking a bit further. For example, in code such as:

```
1  if  (x == 42) then begin
2      do(something);
3  end;
```

the 'do(something);' is a child of 'begin' but in this case it is not indented relative to it. Instead, it is indented relative to 'if' (with an offset of 4, here), which is the parent of 'begin'. The same applies here to 'end'. The choice of the sibling or parent relative to which the code should be indented (and with which offset) is a question of taste and each language and style has its own rules. For example, some styles prefer:

```
1  longfunctionname(argument1, argument2,
2                   argument3, argument4);
```

where 'argument3' is indented relative to its sibling 'argument1', while others prefer:

```
1  longfunctionname(argument1, argument2,
2    argument3, argument4);
```

where 'argument3' is indented relative to its parent, i. e. the beginning of the function call. What is considered as the immediate "parent" actually depends on how fine a granularity one is considering: in the present example, one could instead consider that the immediate parent of 'argument3' is a '..,..' node, which itself has as immediate parent a '(..)' node, whose immediate parent is a function call node.

Indentation rules share a fair bit of commonality between different languages, but each language (and style) has its own particular details.

### 2.2 The limits of automatism

While most source code lines can be fully auto-indented, there are cases where this is either impossible or undesirable. For this reason, editors like Emacs consider auto-indentation as an interactive editing operation that needs to be triggered explicitly by the programmer. For the same reason, auto-indentation only modifies a particular part of the text, specified by the user (typically a single line or a single function).

In languages such as Python, Haskell, and YAML, where whitespace (or more specifically indentation) is significant, indentation cannot be fully automated. In such languages, several indentation choices can be equally valid and result in different syntax trees, so only the human can decide which one is correct. In those languages, the auto-indentation rules will typically return a set of different possible indentations and let the user choose between those.

In other languages, this problem does not appear, but there can still be cases where the user will want to override the auto-indentation. This can happen simply because the indentation rules implemented do not correspond exactly to the coding style the programmer follows, or because of an error in the auto-indentation code, or because the code is sufficiently special that it requires ad-hoc indentation patterns. For example, in the following code:

```
1  compile(mkif(cond1, case1,
2               mkif(cond2, case2,
3                    mkif(cond3, case3, …)  )));
```

where 'mkif' is assumed to be a function that builds some sort of 'if' node in an internal syntax tree, the programmer may prefer to indent it as:

```
1  compile(mkif(cond1, case1,
2          mkif(cond2, case2,
3          mkif(cond3, case3, ...)  )));
```

in which case the source code indentation does not reflect the syntactic structure of the source code, but rather the syntactic structure of the code built at run time.

These limits to auto-indentation mean that the user may tolerate occasional errors from the auto-indentation code. It in turn gives a lot of design room for the indentation code, where very popular languages will usually use a more refined algorithm, whereas esoteric languages will be limited to a very crude auto-indentation code that will more often make poor choices.

It also means that the indentation code should strive to obey previous choices that the user made. For example if the user wants to indent its code in the following unconventional way:

```
1  longfunctionname(argument1, argument2,
2                   argument3,
3                   argument4);
```

while the user should not be surprised if the auto-indenter tries to align 'argument3' with 'argument1', it would be reasonable for them to expect that 'argument4' stays put by simply aligning it with its nearest sibling rather than with the earlier 'argument1'.

### 2.3 Finding siblings and parents

Indentation rules typically require finding siblings and parents since indentation is done relative to their position. This can be done in a number of ways.

**Full parse**   Maybe the most obvious way is to take an existing parser for the source language and perform a full parse of the source code. In order for it to be fast enough one may want to use an incremental parser [7, 17, 15, 16]. Of course, this parser needs to carefully preserve the source code location corresponding to each node of the syntax tree, since the indentation rules will need to go back and forth between the source code and the syntax tree.

This is an attractive approach when the text editor can be tightly integrated with an existing parser, as is the case usually in IDEs where the editor already has access not just to a parser but to a complete language implementation. But it can also suffer from limitations for example when preprocessing or metaprogramming makes it so that elements of the source code are absent from the syntax tree.

**Ad-hoc backward parsing**   In the context of Emacs, reusing an existing parser is often impractical, so most *major modes* (packages that support a particular source language) use instead some kind of approximate and ad-hoc backward parse of the source code, tightly integrated with the indentation rules. For example, when computing the indentation of a line that starts with the keyword 'else', the code will simply search backward for the keyword 'if'.

The main reason why this approach is used is that it can be implemented incrementally, driven by actual cases: the auto-indentation code will start very small, only handling a few important or easy cases (and hence mis-indenting most other cases) and grow over time as needed, usually driven by a growing number of bug reports from a growing number of users.

**Ad-hoc forward parsing**   A kind of middle ground between the two is to perform an ad-hoc approximate parse of the code forward, starting from some "safe place". This "safe place" needs to be far enough that it comes before the relevant sublings and parents relative to which we will want to indent, but close enough that the parsing will be quick. Finding a good "safe place" often relies on clues in the code, such as a piece of code indented in column zero (i. e., not indented at all).

The parsing itself proceeds by moving forward looking for elements such as important keywords, open or close parentheses or braces, and ignoring the rest. This ends up building an approximation of the syntax tree near line $N + 1$.

Like a backward parse, it can be implemented incrementally and driven by concrete cases, progressively refining the information collected in the forward parse as the code matures, but it typically needs a bit more upfront effort: not only does it need a good heuristic to choose its "safe places" but it also needs to parse more code than a backward parse, so it will be thrown off more often by syntax errors or syntactic constructs it does not (yet) understand.

## 2.4  Difficulties with full parsing

The difficulties faced by auto-indentation codes depend on the approach taken. For algorithms relying on ad-hoc parsing, the main problems are linked to the fact that the approach is unprincipled, with different techniques used for different languages or different elements of a language, and potentially complex interactions between them. While the initial effort to get something useful for a given user is fairly low, the effort to scale it to something acceptable to most users is much higher, and the progressive growth tends to be fairly anarchic, leading to code that is very hard to understand and maintain.

Using a full parser takes more effort upfront (unless you can reuse an existing parser) but makes the approach more principled and helps bring structure, making the code much more maintainable. Yet, this approach comes with its own hurdles, even if you start with an existing parser. For example, in many cases, the original parser is designed for a tool such as a compiler and hence focuses on parsing a complete file at a time, so it might require adjustments such as adding some kind of cache mechanism, or making the algorithm incremental in order to avoid performance problems.

**Syntax errors**   Other adjustments will usually be needed to account for syntax errors: when parsing code for a compiler one can presume that the code is syntactically correct and errors should just be detected and reported accurately (with some effort put into trying to recover from syntax errors so as to be able to report more than one error without drowning the user in an avalanche of syntax errors). In the context of auto-indentation in contrast, since it is used as editing is happening, it is commonplace for the code to be syntactically incorrect, so it is important for the parser to have a very good error recovery in order for the indentation not to be unduly affected by syntax errors [11, 8, 3].

Ad-hoc approaches, are not significantly affected by this problem because they work locally, and because the auto-indenter does not need to *detect* syntax errors, contrary to a compiler for example, so it is free to take it for granted that the code is syntactically correct when convenient. For example, it can blindly skip over matched parentheses without looking inside to see if the code therein is valid. Similarly, when looking for the 'if' corresponding to a particular 'else' it does not need to worry about what to do when there is no such 'if': pretty much any answer will do.

**Sadistic syntax**   Even in the absence of syntax errors, a full parser can encounter situations that are difficult to handle. For example, in a language like LaTeX, we usually want to align '\end{FOO}' with the matching '\begin{FOO}', but in reality the language does not guarantee that they come in matched pairs in the source code because these are really commands implemented on top of TeX, so they have to be properly nested during execution but not in the source code.

Similarly, in C the preprocessor introduces a disconnect between the parse tree and the source code since the parse tree is normally built from the output of the preprocessor rather than from the source code itself. For example, consider:

```
1  #ifdef ATOMIC
2    { int x = 42;
3      atomic {
4  #else
5    {
6  #endif
```

After this code, are we within 1, 2, or 3 levels of nested braces?

Another kind of problem with preprocessors is that one might need to indent code that does not exist in the syntax tree. For example, in cases such as:

```
1  #define ANNOTATE(kw) kw
2  …
3    ANNOTATE(if) (test)
4      doit() ;
5    else
6      dont();
```

we would ideally want to align 'else' not with its matching 'if' but with 'ANNOTATE' which would normally not even exist in the syntax tree because it has been removed by macroexpansion before parsing takes place.

**Unknown syntax**   Yet another problem is when the code uses a version of the language not known to the parser. For example if the 'ATOMIC' above reflects the availability of a feature not implemented in the compiler integrated with the editor, the syntax tree will likely reflect the '`#else`' branch of the code, but the programmer might still want to edit the other branch and ask to indent it even though its code might be considered as syntactically incorrect by our parser because it uses constructs unknown to it.

For all these reasons, auto-indentation often looks more like an art than a science, and no matter which approach one takes, syntax errors need to be treated as a normal occurrence and should still result in sane behavior, especially in unrelated parts of the source code that have no good reason to be affected.

### 2.5 Parsing backward

Parsing code backward correctly is harder than forward: the syntax of most programming languages is designed so it can be parsed not only without ambiguity but without even backtracking, but that's only true when parsing forward.

To take a simple example, when parsing backward a C file, if one encounters a '`*/`' which marks the end of a comment, one cannot find the comment's beginning by simply looking for the nearest '`/*`' since this one might itself be within the comment:

```
1  /* This is  /* a single  comment. */
```

Similarly when encountering an LF character (which marks the end of a line), this may be the end of a '`//`' comment, so one needs to look for a '`//`' in the corresponding line. If one is found, again it's possible that this is still not the beginning of the comment, but it's also possible that there's no comment after all, as in:

```
1  mystring = "a  //  b";
```

And of course, if the language allows strings to span several lines, this '`//`' may yet be a comment starter, depending on the code on previous lines.

These kinds of problems do not only occur in the lexical analysis of comments: similar problems occur for the actual syntax analysis, where, when parsing backward, it is often the case that the parsing rules to apply depend on some context that comes before in the text, so when parsing backward it may be necessary to find that context before we can decide how to parse a chunk of source code. Take for example the following code in OCaml:

```
1  let  x = a,  b in x = a,  b
```

This should be parsed as '`let  (x = (a, b)) in (( x = a),  b)`', so when parsing backward a chunk like '`x = a,  b`', one needs to look "ahead" at the preceding code to decide whether to parse it as the expression '`(x = a),  b`' or as the definition '`x = (a,  b)`'.

## 3 Overview

### 3.1 Emacs

Emacs is a text editor implemented in part in C and in part in its own extension language called Emacs Lisp (often shortened to ELisp) [10, 12]. The text on which the user operates is kept in objects called *buffers* where each buffer typically holds the content of a file. At any given moment, one of the buffers is designated as the *current-buffer* and the *point* is a particular position between two characters within this buffer which acts as "the place where we are currently". ELisp primitives like goto-char let you go elsewhere in the buffer, by changing *point*. Most ELisp primitives that act on buffers, such as deleting some characters or inserting a string, operate on the *current-buffer* at *point*.

Every buffer is associated with a *major mode*, which can adjust the behavior of Emacs to the particular content of the buffer. For example, for a buffer that contains Javascript code, the major mode will usually be set to js-mode. Major modes are implemented in ELisp and can provide mode-specific key-bindings, as well as set *buffer-local* properties which affect the behavior of the generic code in this particular buffer. For example, js-mode sets the comment-start property to "//" to indicate the string that should be used by the generic comment-region function when inserting comment delimiters.

## 3.2 SMIE

SMIE is an indentation engine implemented in ELisp and introduced in Emacs-23.3. It includes a parser that can work both forward and backward, but its auto-indentation algorithm is fundamentally based on backward parsing. This choice was made for several reasons:

- Of the three approaches, parsing backward is the one that is best protected from syntax errors: it naturally stops parsing as soon as it finds the parent relative to which the code needs to be indented, so it is inherently unaffected by syntax errors that might be found further away.
- For many (if not most) languages, the *major mode* that provides support for it in Emacs is not written by an implementer of the language but by a user whose goal is to get something that's "good enough" for personal use as quickly as possible. Having to perform a full parse building a syntax tree as a first step would likely discourage them.
- Backward parsing was the most commonly used approach in pre-existing auto-indenters for the Emacs text editor, so there was a lot of experience from which to draw inspiration.

Emacs's pre-existing backward auto-indenters are very difficult to maintain, because the parsing is done in a completely ad-hoc way (typically involving regular expression searches), so each one uses a different mix of different techniques often intertwined in non-trivial ways, with some amount of redundancy added for good measure, and all of it made yet more intricate by having the indentation rules and the actual parsing deeply interconnected. It is often described as "dark magic" where "dark" should be understood as a euphemism for "obscure".

SMIE can be seen as an attempt to make Emacs's auto-indenters more principled and maintainable as well as make it easier for a potential contributor to start writing such an auto-indenter. The goal of SMIE is to provide an infrastructure that makes incremental improvements easy: not only it is easy to get a first cut at an auto-indenter for a new language, but it is also easy to grow this solution to something more robust, without having to restart from scratch along the way.

## 3.3 General structure of SMIE

SMIE divides the problem of automatic indentation into 4 layers, each of which major modes will usually need to configure for their particular use:

- *Syntax-table*: The bottom layer concentrates on distinguishing comments and strings from other parts of the code. This layer is provided by Emacs's built-in system of *syntax-tables*. It relies on a very simple parser implemented in C code, supplemented with a straightforward caching mechanism implemented in Elisp.
- *Lexer*: The second layer divides the remaining part of the code into tokens. This layer is implemented in an Elisp function provided by the major mode. It takes no argument and fetches the next token. It can presume that it is called *between* tokens.
- *Parser*: The third layer is the parser, which can work forward as well as backward and can start parsing a sub-expression from any position between tokens. It relies on a grammar provided by the major mode.
- *Indentation rules*: The top layer is the indentation code itself, which uses the parser to find siblings and parents and relies on an Elisp function provided by the major mode to know with which ones to align and with what offset.

## 3.4 The indentation rules

The main entry point of the indenter is the function `smie-indent-calculate` which computes the column where the first char of a particular lexical element (token) should be placed, which is called the indentation column, or just indentation, of that token. It distinguishes 3 different cases, depending on whether the token is a *keyword*, or is right after a *keyword*, or whether neither applies. All tokens which appear as terminals in the grammar of the parser are considered as *keywords*: 'let' and 'in' but also open parentheses, semi-colons, ...). It then consults the indentation rules provided by the major mode to decide whether to look for a sibling or a parent relative to which it should be indented and uses the parser to find them by parsing the code backward from the position of interest.

## 3.5 The parser

The design of SMIE revolves around parsing the code backward in order to find the siblings and parents of a given piece of code. Yet, it is also sometimes necessary or at least useful to parse the code forward, either as part of the indentation rules, or when using the parser for other purposes such as to let the user navigate the code conveniently.

For this reason, the parser of SMIE is designed to work forward as well as backward. We know of two classes of parsing techniques which can use the same grammar to parse both forward and backward: the techniques like GLR (Generalized LR) [13] which can handle any context free grammar, and the techniques based on OPG (operator precedence grammar) [5].

Those two classes sit at opposite ends of the spectrum, where GLR parsers can handle much more complex grammars than is usually needed to parse programming languages, whereas OPG parsers are too weak to parse correctly all but the (syntactically) simplest programming languages. The reason why GLR can work both ways is that it is so strong that it can overcome the difficulty, and conversely OPG can work both ways because it is so weak that it cannot take advantage of any particular direction.

In SMIE, we decided to go with the weaker option.

OPG parsers can work in both directions because they only handle grammars which we could call "*strongly* context free": in our OCaml example, while the grammar is context free, it can parse 'x = a, b' either as the definition 'x = (a, b)' or as the expression '(x = a), b' and this ambiguity is resolved by the context, depending on whether an expression or a definition is expected. In an OPG grammar, such an ambiguity is disallowed: either 'x = a, b' is always parsed as 'x = (a, b)' or it's always parsed as '(x = a), b', regardless of the context.

This is a major weakness that SMIE needs to address to be viable (see below), but it has a flip side. It means that not only OPG parsers can work in both directions, but they can also work locally, since they do not need to parse the rest of the code just to get the context which determines *how* to parse the local code. This is a key benefit for us: SMIE does not need to perform a full parse of the text; it can limit its parsing to the local piece of code that is under consideration. This is a source of significant simplification, since it eliminates most performance concerns, making it unnecessary to use incremental parsing or to keep a cache of previous partial parses. It also makes the system naturally robust against syntax errors in distant unrelated pieces of code.

### 3.6 The lexer

Parsing in a compiler is typically divided into a lexer and a parser, where the lexer processes the input *stream* of characters with some kind of finite automaton. This imposes fairly severe constraints which usually drive the choice of what to handle in the lexer and what to handle in the parser. More specifically, it is common to delay some of the work to the parser (or to let the parser give more context to the lexer as in [14]) because it can't conveniently fit within the constraints of stream processing or within the constraints imposed by a finite automaton.

In SMIE, in contrast, the main constraint in the division of labor between the parser and the lexer comes from the use of an OPG parser. To make up for it, the lexer is not limited to a finite automaton, nor is it limited to stream processing: it is free to look at all the surrounding text (including using the SMIE parser recursively) in order to decide what should be the next token.

Concretely, in the case of our OCaml example, in order to decide how to parse 'x = a, b', since the OPG parser cannot have two different parsing choices for the same sequence of tokens, we have to arrange for the lexer to return different tokens depending on the surrounding context. In Tuareg-mode (a major mode for Ocaml files), this is done by returning different tokens for the '=', depending on whether it corresponds to an equality test or a definition. The distinction is done by ad-hoc code searching the text before the '=' character for tell tale signs, such as encountering an 'in' or a 'let' token.

### 3.7 The syntax tables

The syntax tables are a primitive parsing system that has been part of Emacs since its early life. They map every character to a few categories, basically indicating if that character is a kind of open parenthesis, close parenthesis, string delimiter, escape character, whitespace, comment opener, comment closer, or can be part of an identifier. They can also handle multi-character comment delimiters via some ad-hoc extensions.

Using this information, they offer functions to skip over a balanced pair of parentheses or an identifier or a string, and functions to skip over whitespace, including skipping over comments, all of that working equally well forward as backward. Also, since Emacs-20, they offer a function, called parse-partial-sexp, to find the (forward) parsing state at a given position in a *buffer* (the internal representation of the content of a file), which in Emacs-22 was supplemented with the function syntax-ppss which does the same but keeping a cache in order to speed up this computation.

The parsing state is basically limited to counting the parenthesis depth and indicating whether the position is within a string or a comment. This is sufficient to implement good navigation and automatic indentation functionality for languages with a very simple syntax, like Emacs Lisp, but does not scale to most other languages's syntax. Syntax tables can still be useful for other languages but only to a limited extent. SMIE uses it to jump over comments, strings, or balanced pairs of parentheses or braces, as well as to discover whether a given position is within a comment or a string.

## 4 Parsing with OPG

### 4.1 Low-level representation of the grammar

Operator precedence grammars take the form of a precedence table, which is usually represented as a 2-dimensional array indicating for every pair of tokens what is their relative precedence. For example, a precedence table for a small language with multiplication, addition, and let...in... might look like:

|     | *id* | let | in | + | * |
|-----|------|-----|----|----|----|
| *id* |      |     | ⋗ | ⋗ | ⋗ |
| let | ⋖ | ⋖ | ≐ | ⋖ | ⋖ |
| in  | ⋖ | ⋖ |   | ⋖ | ⋖ |
| +   | ⋖ | ⋖ | ⋗ | ⋗ | ⋖ |
| *   | ⋖ | ⋖ | ⋗ | ⋗ | ⋗ |

**SMIE: Weakness is Power!**

The way to read this table, for example for the top-right entry, is: '*id* ⋗ *'  means that *id* binds more tightly than '*' when it is to the left of a '*'. A good way to think about it is that a ⋗ rule will introduce an implicit close parenthesis between the two tokens whereas a ⋖ will introduce an implicit open parenthesis. An ≐ sign indicates that those two tokens go together, as shown above with the entry 'let ≐ in'. The two empty spots in the first line indicate that *id* cannot be followed by another *id* or by a 'let'. The bottom-right entry '* ⋗ *' expresses the fact that '*' associates to the left.

Instead of using such a 2-dimensional precedence table, SMIE uses what is sometimes called *precedence functions*: we use a simple map which assigns to every token a left and a right precedence level. For example, 'let ≐ in' is represented by the fact that the right precedence level of 'let' is equal to the left precedence level of 'in', whereas '+ ⋗ +' is represented by the fact that the right precedence level of '+' is larger than its left precedence level. So the previous 2-dimensional table could become:

| Token | *id* | let | in | + | * |
|---|---|---|---|---|---|
| left  | 7 | 6 | 0 | 2 | 4 |
| right | 7 | 0 | 1 | 3 | 5 |

This places an additional restriction on the grammars we can handle: it disallows empty entries like the one at the top-left of the table, and it forces the ⋖ and ⋗ relations to be transitive. The choice of using precedence functions instead of 2-dimensional tables in SMIE was largely arbitrary to keep the tables smaller, but can be justified as follows: since empty entries only serve to express that some forms are syntactically invalid, SMIE cannot make use of them, and filling those entries by assuming the relation to be transitive ensures a more sane handling of such syntax errors anyway. In practice we have found that the difference is rarely noticed, apparently in agreement with the general consensus [1].

**Associativity**  In SMIE, infix keywords like '+' or ';' are usually given the same left and right precedence, which tells SMIE that they are so-called (syntactically) *associative*: while the language's real syntax might technically consider 'a + b + c' as '(a + b) + c' and this may sometimes matter because of rounding errors, the code's indentation does not care to reflect this detailed structure. The end result is usually not affected, but using the same precedence on both sides tells SMIE that 'a', 'b', and 'c' are all siblings, so when asked to indent 'c' it may decide to align it with 'b' without having to look at 'a', which is not only more efficient but can avoid mis-indentation if there's a syntax error between the two. This is basically the same notion of associativity as that used by Wagner [16] to turn potentially long linked lists of non-terminals into shallower trees.

### 4.2  Sequences of non-terminals

SMIE grammars differ from typical OPG grammars in that they can recognize some repetitions of non-terminals only separated by whitespace. This is done automatically, without any explicit entry in the precedence table:

- The table does not record the rules for identifiers: any token not found in the table is assumed to be an identifier and is treated as an atomic element that does not bind at all on either side. As mentioned earlier, this is also used to distinguish *keywords* from other tokens.
- Instead of being a number, the left or right precedence level of a keyword can be *nil*, indicating that this keyword does not bind at all on this side. For example, the left precedence of '(' and the right precedence of ')' are typically *nil*. This way, 'a(b)c' is not parsed as a single node '..(..)..' with the three children 'a', 'b', and 'c', but as a sequence of 3 nodes, the middle node being a '(...)' with a single child.
- The parser automatically accepts consecutive sequences of non-terminals. For example, 'a + b c ¬d' will be parsed as a '+' node with two children, the right child being a sequence of three elements: the identifiers 'b' and 'c' and the node '¬' which has as single child the identifier 'd' (assuming a precedence table with entries for '+' and '¬', where the left precedence of '¬' is *nil*).

This change is important since it is common for programming languages to accept sequences where the elements are not separated by special tokens. This is the case for curried function calls in most functional programming languages, for example.

## 4.3 Other representations of the grammar

For a source language like Prolog, it is straightforward to build a precedence function from the language's own precedence table, but for most other languages it is rather inconvenient.

For this reason, SMIE lets you provide the grammar as a 2-dimensional precedence table, where each entry can be one of $<, =, >$, or empty. It then takes this table as a set of constraints and finds an assignment of precedence levels which satisfies those constraints (or signals an error if it can't be done).

While this is sometimes a bit more convenient, it is still rather cumbersome to use, so SMIE additionally allows you to provide the grammar as a sort of BNF specification. Each production rule in that specification is restricted to have at least one terminal between each pair of non-terminals. This BNF grammar can then be passed to an SMIE function to generate a corresponding 2-dimensional precedence table.

To go along with that, there is also a function that builds such a 2-dimensional precedence table from an ordered list of un-numbered precedence levels where each level contains a set of keywords and is annotated as either left-associative, right-associative, non-associative, or *associative* (see section 4.1).

Finally, SMIE can combine several 2-dimensional precedence tables into one. This is typically used in the following way: the language's grammar is described as a set of BNF rules plus a set of precedence levels (many languages come with such a list of infix operators segregated into a dozen or so precedence levels), the two are then combined into a single 2-dimensional precedence table which is finally converted to the low-level simple precedence table.

### 4.4 Navigation

SMIE does not build parse trees. Instead, the only functionality its parser offers is that of navigation: its core function is smie-next-sexp which jumps over the next sub-expression, either forward or backward. The caller can also provide it with a keyword that needs to be matched. For example, one can call smie-next-sexp backward, providing it with the keyword 'then', in which case it will skip over all the text that could appear to the left of 'then', presumably stopping at the matching 'if'. Its return value indicates the keyword against which it bumped, when applicable.

For example, with the following text:

```
1  (x + b * c d
```

and with *point* at the end of the line, if we call smie-next-sexp backward without any keyword, it will just jump over 'd' and return that it did not bump against any keyword. If instead we call it with a keyword '*' it will jump over 'c d' and return that it bumped into another '*' (this assumes that '*' is marked as associative rather than left-associative). If instead we call it with a keyword '+' it will jump over 'b * c d' and return that it bumped into another '+'. And finally if we call it with a keyword ')' it will jump over the whole line and return that it did not bump into any keyword.

One can also call smie-next-sexp with a special argument 'halfsexp' which tells it to skip over the other side of the very next keyword. If we take the previous example, with *point* between '*' and 'c', and we call smie-next-sexp backward with 'halfsexp', it will jump over 'b *' and return that it bumped into '+': it first read the keyword '*' and then jumped over its left hand side argument. This is often very handy to jump directly to the sibling.

## 5 Indentation rules

The indentation algorithm looks at the immediately preceding and following tokens to see if we're indenting a keyword, or indenting code immediately following a keyword or neither. It defers the details of each decision to a function provided by the major mode. SMIE calls that indentation-rules function with a description of the case at hand and usually expects in return the indentation offset that should be used.

If we're indenting a keyword, we will usually want to jump backward over its left-hand side to find either the keyword of its parent node (e. g. we started with the keyword '*' and we found '+') or a matching keyword (e. g. we started with the keyword 'then' and we found 'if'). We then usually align the keyword either with its matching keyword, or with the text that immediately follows the parent keyword. Before doing that, we call the major mode's indentation rules function so it can choose a different indentation.

If we're indenting right after a keyword, we will usually indent the text relative to that keyword, with a standard offset. But here again, we first call the major mode's indentation rules function so it can either return a different offset or choose a completely different indentation.

Finally, when we're neither after nor before a keyword, it means we're inside a sequence of non-terminals, between two such non-terminals. By default, SMIE presumes that this sequence should be indented like a curried function call, where the second non-terminal should be indented with an offset relative to the first, and subsequent ones should be aligned with the second:

```
thefunction
    arg1
    (arg2 expression)
    arg3
```

but it will first look at the keyword before the first non-terminal and then consult the major mode's indentation rules function, in case these non-terminals should be all aligned at the same indentation.

## 5.1 Virtual indentation

When we say that code should be aligned with another part of the code, it sounds simpler than it really is. For example, in a language like C we usually say that we want to indent the content of '{...}' relative to the opening brace with an offset of, say, 4:

```
int function (int arg)
{
    dosomething();
}
```

But that does not mean we want an indentation like:

```
int function (int arg) {
                       dosomething();
                       }
```

SMIE handles this via the notion of *virtual indentation*: when we indent relative to the opening brace, we first compute the virtual indentation of that brace. If the brace is at the beginning of the line, then its virtual indentation is its current indentation, but if it is elsewhere on the line, we recursively call smie-indent-calculate to find the column at which this brace "would be" indented.

So the indentation rules need to be able to compute the intended indentation of any position in the buffer, whether at the beginning of line or not. And the indentation rules can explain the special treatment of the opening brace by noticing that it is what we call *hanging*: it is placed at the end of a line and is not at the same time at the beginning of a line (i. e., it's not alone on its line).

In the above example, the virtual indentation of the opening brace would then presumably be the column of the first 'int', so both the 'dosomething();' and the closing brace would then be aligned relative to that part of the code:

```
int function (int arg) {
    dosomething();
}
```

▪ **Listing 1**   Sample Relax-NG Compact file recettes.rnc

```
1  datatypes xsd = "http://www.w3.org/2001/XMLSchema-datatypes"
2
3  start  = element recettes { recettes }
4
5  recettes = recette+ |
6             element group {
7                attribute  nom { string },
8                recettes
9             }
10
11 recette = element recette {
12            attribute  nom { string },
13            attribute  photo { xsd:anyURI }? ,
14            ingredients,
15            etapes
16         }
```

The concept of virtual indentation simplifies the indentation rules as can be seen here since changing the virtual indentation of the opening brace corrects two cases in a single-shot: the indentation of the closing brace and the indentation of the body.

The effect can be even more important when the virtual indentation ends up applied recursively. For example, in Standard ML the syntax for anonymous functions is 'fn *arg* => *exp*' and '*exp*' should be indented relative to 'fn', but in code like:

```
1  fn x => fn y => fn z =>
2              x + y + z
```

we usually want to indent the body of the curried function much less deeply. This can be done very naturally by simply saying that when we compute the virtual indentation of 'fn' we look if it happens to immediately follow a '=>' in which case we want to align it with its parent (which here will be the previous 'fn'). This will then recursively compute the virtual indentation of the previous 'fn' until reaching the first, resulting in the desired indentation:

```
1  fn x => fn y => fn z =>
2    x + y + z
```

## 6   Example

Let's try to write the indentation support for rnc-mode which provides support for the Relax-NG Compact syntax, an EBNF-like syntax to specify XML schemas. Listing 1 shows part of a sample file recettes.rnc

### 6.1 Syntax-table for RNC

The first step is to specify the syntax-table, which will also be used by other parts of Emacs. It might look like the following:

```
1  (defconst rnc-mode-syntax-table
2    (let (( st (make-syntax-table)))
3      (modify-syntax-entry ?\{ "(}" st)
4      (modify-syntax-entry ?\} "){" st)
5      (modify-syntax-entry ?\" "\"" st)
6      (modify-syntax-entry ?# "<"  st)
7      (modify-syntax-entry ?\n ">" st)
8      (modify-syntax-entry ?:  "_"  st)
9      st))
```

The syntax is a bit arcane, but all this is doing is telling Emacs that braces are matched, that double quotes delimit strings, that '#' starts a comment and linefeed ends it, and that ':' should be in the syntax category of characters that can occur within an identifier rather than in the category of punctuation characters.

The first three calls to `modify-syntax-entry` are actually redundant because `rnc-mode` already inherits the default syntax-table settings for programming major modes which declares the parentheses and braces as paired characters and it also declares '"' as a string delimiter.

At this point, if we setup SMIE without any extra information, it will make use of the information available in the syntax table (mostly the fact that braces come in matched pairs) to provide a minimal amount of indentation support. Unsurprisingly, it will indent most of the example in listing 1 wrong.

### 6.2 Grammar for RNC

Some people may be tempted to take an existing BNF grammar and plug it in, but this is a recipe for a big disappointment: SMIE's grammars may look like BNF grammars, but they're fairly different because a grammar designed for a typical LR parser will usually work poorly with an OPG parser. Also, the indentation rules require that the abstract syntax tree match closely the BNF derivation tree, because they basically interpret the derivation tree as an abstract syntax. For these reasons, it is preferable to build the BNF grammar by hand, and it is also better to do it bit by bit.

If we look at the sample file, we see that the file is structured as a sequence of declarations '*id = pattern*'. But there is no token that separates those declarations, so our parser won't be able to handle that: while it can handle sequences of non-terminals, this only applies when each element in the sequence is itself naturally delimited, either because it is a single identifier or because it is wrapped inside parentheses or similar tokens with a *nil* precedence level on the outside.

When something cannot be done in the parser, we can push it to the lexer: we will arrange for the lexer to recognize the whitespace between each declaration and return the pseudo keyword ' ; ' for it.

The RNC patterns can be combined with ',' or with '|' and they can take postfix multiplicity annotations such as '?' and '+' and can be of the form 'element *id* { *pattern* }' or 'attribute *id* { *att-spec* }'.

Based on those descriptions, one could write the following SMIE grammar:

```
 1  (defconst rnc-smie-grammar
 2    (smie-prec2->grammar
 3     (smie-bnf->prec2
 4      '( (id)
 5         (args (id)    ("{"  pattern "}"))
 6         (decls (id   "=" pattern)
 7                (decls " ; " decls))
 8         (pattern ("element" args)
 9                  ("attribute" args)
10                  (pattern ","  pattern)
11                  (pattern "|"  pattern)
12                  (pattern "?")
13                  (pattern "+")))
14      ;; Resolve precedence ambiguities.
15      '((assoc " ; "))
16      '((assoc ","  "|")  (nonassoc "?" "+")))))
```

Notice that the '*id*' category is declared without any rule, because on the one hand we don't want to have terminals to represent mere identifiers and on the other, a simple identifier (as well as any subexpression wrapped within matched parentheses or braces) is always allowed to appear anywhere. Along the same lines, for '*args*' we take advantage that repetition of non-terminals are automatically allowed, so we basically accept 'element' followed by any number of '*id*' or '{ *pattern* }'.

Notice also that the BNF part of our grammar specifies rules for infix keywords like '|' which lead to ambiguities. Rather than resolve them by stratifying '*pattern*', SMIE lets us provide auxiliary precedence rules after the BNF, which are used specifically to resolve those ambiguities. In this example, those precedence rules are divided into 2 sets, so as to spare us the need to decide the relative precedence between ' ; ' and the other tokens.

If we try to use SMIE with just this grammar and nothing more, it does not work noticeably better than before: it also gets most of the lines wrong, except for 5. But this time those 5 correct lines are not an accident. Earlier 'ingredients' was correctly aligned with 'attribute', but only because it happened to be at the beginning of a line (',', '{', and 'photo' were incorrectly considered a siblings as well). This time it is aligned because SMIE understands the syntax of ',' and hence knows that the previous 'attribute' is its true sibling.

**6.3 Lexer for RNC**

The default lexer in SMIE uses the syntax tables to divide the text into tokens that can be either identifiers, punctuation, or open or close parenthesis or brace. This actually works well for us with two exceptions: the first is the already mentioned need to fabricate ' ; ' pseudo keywords when skipping over the whitespace between declarations and the second is that the default lexer considers any sequence of consecutive punctuation characters as a single punctuation token, whereas we need to treat '?,' as two separate tokens.

We can easily recognize the whitespace between declarations, since it is always followed by 'id =', so the code of the lexer can look like:

```
 1  (defun rnc-smie-backward-token ()
 2    (let ((start (point)))
 3      (forward-comment (- (point)))
 4      (if (and (< (point) start)
 5               (let ((pos (point)))
 6                 (goto-char start)
 7                 (prog1
 8                     (looking-at "\\(?:\\ s_\\|\\sw\\)+[ \t\n]*=")
 9                   (goto-char pos))))
10          " ; "
11        (if (looking-back "\\s." (1- (point)))
12            (buffer-substring-no-properties
13             (point)
14             (progn (forward-char -1)
15                    (point)))
16          (smie-default-backward-token)))))
```

When we pass this lexer to SMIE, we start getting an indentation that makes sense: it's not quite right at places, but it is already quite usable.

### 6.4 Indentation rules for RNC

With the lexer in place, SMIE now gets almost all lines right in our sample file, except for the two closing braces and the first line of the body of the elements (the subsequent lines in those bodies are all indented correctly, aligned with the first line). All four are actually due to the same problem in the calculation of the virtual indentation of the opening braces: in 'element id {...}', since 'element' is a keyword and the rest isn't, SMIE parses it as a node 'element' with one child which is a sequence of length two, which gets indented as a function call. More specifically, SMIE here indents the '{...}' block by aligning it with the 'id', adding an offset which defaults to 4 columns. Instead, we want to align the '{' with 'element', at least when the '{' starts the block of an 'element'. We can do that with the following indentation rule function:

```
1  (defun rnc-smie-rules (kind token)
2    (pcase (cons kind token)
3      (`(:before . "{")
4       (save-excursion
5         (rnc-smie-backward-token)
6         (when (member (rnc-smie-backward-token)
7                       '("" "element" "attribute"))
8           `(column . ,(smie-indent-virtual)))
9         )))
```

Of course, once this is done, the coder will try another RNC file and discover that its grammar only covers a subset of the language and that the indentation rules aren't quite right yet for some use cases, so the above is only a starting point. Yet,

**SMIE: Weakness is Power!**

■ **Table 1** Known SMIE indenters and their size in kB

| Language | Origin | Syntax-table | Grammar | Lexer | Rules | Total |
|---|---|---|---|---|---|---|
| Coq | other | 1.1 | 5.0 | 30.8 | 12.4 | 49.3 |
| OCaml | other | 3.5 | 11.2 | 15.3 | 14.8 | 44.8 |
| Haskell | other | 6.2 | 1.4 | 14.9 | 16.3 | 38.8 |
| Elixir | other | 6.5 | 2.1 | 6.1 | 19.4 | 34.2 |
| sh | Emacs | 11.3 | 1.3 | 6.1 | 6.9 | 25.7 |
| C | ELPA | 2.2 | 3.7 | 6.8 | 11.4 | 24.1 |
| Ruby | Emacs | 9.2 | 2.4 | 6.8 | 4.2 | 22.5 |
| rc(Plan9) | Emacs | 11.3 | 0.7 | 4.0 | 1.1 | 17.0 |
| SML | ELPA | 0.5 | 4.5 | 3.5 | 3.8 | 12.4 |
| Octave | Emacs | 2.8 | 4.0 | 3.4 | 1.7 | 11.8 |
| Smalltalk | ELPA | 3.8 | 1.8 | 2.9 | 3.1 | 11.6 |
| Modula-2 | Emacs | 0.7 | 4.3 | 2.2 | 1.0 | 8.2 |
| Beluga | other | 0.8 | 3.4 | 2.2 | 1.6 | 8.0 |
| Prolog | Emacs | 2.0 | 1.6 | 1.1 | 2.5 | 7.2 |
| GLE | ELPA | 1.2 | 0.9 | 3.8 | 0.3 | 6.2 |
| CSS | Emacs | 1.5 | 0.3 | 1.9 | 0.6 | 4.4 |
| Wolfram | other | 2.0 | 1.5 | 0.0 | 0.7 | 4.1 |
| GAP | other | 0.8 | 1.5 | 0.0 | 1.7 | 4.0 |
| Postscript | Emacs | 3.2 | 0.0 | 0.0 | 0.2 | 3.4 |
| RNC | ELPA | 0.3 | 0.9 | 1.0 | 1.0 | 3.3 |
| DTS(Linux) | ELPA | 1.0 | 0.7 | 0.0 | 0.7 | 2.4 |
| Sed | ELPA | 1.8 | 0.0 | 0.0 | 0.1 | 1.9 |
| OPAM | other | 0.4 | 1.0 | 0.0 | 0.3 | 1.7 |
| JBuild | other | 0.5 | 0.1 | 0.0 | 0.7 | 1.3 |
| JSON | ELPA | 0.4 | 0.1 | 0.0 | 0.2 | 0.7 |

the important property here is that the structure makes it fairly easy to add new constructs and new rules incrementally.

## 7 Experience

Since its introduction in Emacs-23.3 in 2011, SMIE has been used in a growing number of major modes. Table 1 shows the 25 SMIE indenters we have found: 8 bundled with Emacs, 8 in the GNU ELPA repository hosted by Emacs maintainers, and 9 more in the wild. Half of those indenters were mostly developed by the original author of SMIE.

This shows that SMIE can handle a large variety of syntaxes, well beyond what the limits of OPG might suggest. It also shows that the grammar itself is almost never the larger component, which is a direct consequence of the weakness of the parsing technology. In a sense, the main benefit of the use of OPG is as a wedge between the indentation rules and the lexer, enforcing a cleaner structure than the ad-hoc mess

■ **Table 2** Code size of other approaches relative to SMIE

| Language | Ad-hoc % | Tree-sitter % |
|---|---|---|
| Coq | 61 | — |
| OCaml | 159 | 135 |
| Haskell | 110 | 178 |
| sh | 164 | 58 |
| C | — | 101 |
| Ruby | 162 | 118 |
| SML | 158 | — |
| Octave | 39 | — |
| Prolog | 339 | — |
| CSS | — | 194 |
| JSON | — | 401 |

used before. Some of this "ad-hoc mess" finds its way into the SMIE lexer, of course, but its task is much more focused since it is now separated from the actual parsing and indentation logic, so while many of the "ad-hoc" heuristics are preserved, the "mess" is tremendously reduced.

A point worth noting is also the number of those languages where the total size is very small. This is sometimes the direct result of the underlying simplicity of the language, as is the case for Sed and JSON, and other times it's a reflection of a primitive implementation, as is probably the case for Wolfram and GAP. In either case it shows how SMIE makes it easy to get started. This is reflected also in the reliance on the default lexer in the 8 languages where the lexer's size is 0 kB.

## 7.1 Comparison

Table 2 shows a rough idea of the relative size of the SMIE code for different languages compared with their previous ad-hoc parser as well as with the grammar used by the Tree-sitter tool [2], a state-of-the-art incremental parser based on a GLR parser.

The SMIE indenters all behave at least as well as the ad-hoc ones (it's a necessary condition before replacing the ad-hoc code with the SMIE code), so the first column shows that except for Coq and Octave, the new code is not only better but also smaller. In the case of Coq and Octave, the increase in size is justified by the fact that the ad-hoc code behaved very poorly compared to the new code. We do not show a comparison with the ad-hoc indenter for C for 2 reasons: first, the SMIE indenter for C does not behave as well as the ad-hoc one yet; and second the ad-hoc indenter for C is implemented as part of another indentation engine which weighs in at around 500 kB of code, and includes support for several other languages (including Awk, C++, and Java).

The second column compares the size of the SMIE based parser compared to a state-of-the-art GLR grammar. SMIE does not intend to compete with tools like Tree-sitter: a full parse tree like that maintained by Tree-sitter offers many other benefits which SMIE cannot even dream of. The comparison here intends to look at whether the ability offered by SMIE to grow your parser incrementally comes at the cost of an end result that is larger than if you had paid the cost of a full parser upfront.

As we can see in this table, the SMIE-based parser is still smaller than a GLR-based grammar in all cases, except for one: in the case of the sh language, a full parser is definitely worth the upfront cost. If we look at the breakdown of the size of SMIE's sh parser in table 1 we see that the culprit is the syntax-tables, which are indeed very poorly matched to the structure of the sh language, so there is still hope that an approach like that of SMIE but with a different layer than Emacs's syntax tables at the bottom might suffer a bit less in the comparison.

### 7.2 Why does it work

Experience shows that in practice SMIE's use of an arbitrarily complex lexer is able to overcome the inherent limitations of OPG for a large set of cases. One can even easily show that in theory SMIE can correctly parse any language that can be parsed by a GLR parser: if you take the BNF grammar and surround every non-terminal appearing on a right-hand side with two special new terminals that we could call '(' and ')', you get a BNF grammar that falls in the OPG subset. You can then pair this grammar with a lexer that performs GLR parsing under the hood and then uses the parse tree to emit the original tokens augmented with the necessary synthesized '(' and ')' token.

Real SMIE parsers do not go to such extremes, tho. In practice there are two main sources of restrictions from OPG: the obligation to be what we called "*strongly context free*" and the fact that in production rules two non-terminals always have to be separated by at least one terminal.

The former is naturally solved in the lexer: in the example of OCaml's 'x = a, b', we change the lexer so that when it sees '=' it returns, for example, either 'd=' for a definition or 'e=' for a boolean test depending on the context.

The second restriction can be more severe. To solve these issues, we need to modify the grammar, often in ways which affect the accepted language in minor ways. In practice, the only effect that seems to occur in programming languages is that it disallows kleene-closure repetition, which typically falls into two camps:

- It's a repetition of self-delimited elements (elements are wrapped within something akin to parentheses), which SMIE handles automatically.
- It's a repetition of elements that are marked by special starting or terminating tokens, typically ';'. To handle such cases in SMIE, one needs to change the grammar such that the special tokens become separator tokens.

Sadly, there are cases where we cannot easily change the grammar to use *separators*. The most glaring such cases is with languages using a C-style syntax, where making ';' separate instructions instead of terminating them leads to incorrect parses: e. g., it might misparse 'if (A) B; else {C} D;' as '(if (A) B); (else ({C} D);'. To coerce SMIE into parsing the language more correctly we could make the lexer insert extra synthesized tokens to mark the end of expressions like 'if'. In the current SMIE-based indenter for C we instead use ad-hoc code in the indentation rules which takes into account and compensate for the misparse, which is ugly but expedient.

## 8 Related Work

There is a lot of literature on incremental parsing for structured text editors, such as [7, 17, 15, 16]. They all focus on maintaining a full syntax tree of the edited text.

In [9], Lapalme shows the design of an auto-indenter implemented in Elisp for languages like Haskell where indentation is significant. The basic design of the algorithm he proposes does not really pay attention to the actual grammar of language: it is based on looking at the few previous lines where indentation increases, which are all taken as possible indentation positions. Of course, to make it less naive, it is combined with ad-hoc rules which filter some of those positions based on knowledge of the language's syntax.

## 9 Conclusion

We presented the SMIE indentation engine, which little by little, has been replacing many of the ad-hoc automatic indenters in Emacs. Its most immediate benefit is that it is easy to get started with a simple code and then incrementally grow it to cover additional cases. The less visible benefit is that it inherently degrades gracefully in the presence of syntax errors or incomplete knowledge of the language. It gets those qualities thanks to the use of the very simple OPG parsing strategy whose inherent weakness proves here particularly beneficial, helping guide the structure of the solution to preserve these maintainability and robustness qualities.

Before SMIE, the main problem with Emacs's ad-hoc parsers was that the growing complexity of the code made it humanly impossible to improve the behavior without a complete rewrite and that the only rewrite that seemed to make sense was to write a full parser, a prospect apparently sufficient to discourage all contributors. Systems like Tree-Sitter might still change that, but in the mean time, SMIE seems to argue that worse can sometimes be better [6].

**Acknowledgements** This work was supported by the Natural Sciences and Engineering Research Council of Canada (NSERC) grants № 298311/2012 and RGPIN-2018-06225. Any opinions, findings, and conclusions or recommendations expressed in this material are those of the author and do not necessarily reflect the views of the NSERC.

## References

[1] Alfred V. Aho, Ravi Sethi, and Jeffrey D. Ullman. *Compilers — Principles, Techniques, and Tools*. Addison-Wesley, 1988.

[2] Max Brunsfeld. Tree-sitter - a new parsing system for programming tools. Strange Loop Conference, 2018. Accessed 2020-05-01. URL: https://www.thestrangeloop.com/2018/tree-sitter---a-new-parsing-system-for-programming-tools.html.

[3] Michael G. Burke and Gerald A. Fisher. A practical method for LR and LL syntactic error diagnosis and recovery. *Transactions on Programming Languages and Systems*, 9(2):164–197, 1987. doi:10.1145/22719.22720.

[4] Irène Durand and Robert Strandh. Incremental parsing of Common Lisp code. In *European Lisp Symposium*, pages 16–22, 2018. URL: https://hal.archives-ouvertes.fr/hal-01887230/, doi:10.5281/zenodo.3247555.

[5] Robert W. Floyd. Syntactic analysis and operator precedence. *Journal of the ACM*, 10(3):316–333, July 1963. doi:10.1145/321172.321179.

[6] Richard P. Gabriel. Worse is better. EuroPAL keynote speech, 1989. Accessed 2020-05-01. URL: http://dreamsongs.com/WorseIsBetter.html.

[7] Carlo Ghezzi and Dino Mandrioli. Incremental parsing. *Transactions on Programming Languages and Systems*, 1(1):58–70, 1979. doi:10.1145/357062.357066.

[8] Susan L. Graham, Charles B. Haley, and William N. Joy. Practical LR error recovery. In *Proceedings of the 1979 SIGPLAN Symposium on Compiler Construction*, SIGPLAN '79, pages 168–175, 1979. doi:10.1145/800229.806967.

[9] Guy Lapalme. Dynamic tabbing for automatic indentation with the layout rule. *Journal of Functional Programming*, 8(5):493–502, September 1998. doi:10.1017/S0956796898003098.

[10] Bil Lewis, Dan LaLiberte, Richard Stallman, and GNU Manual Group. *GNU Emacs Lisp Reference Manual*. Free Software Foundation, 3.1 edition, 2018. Accessed 2020-05-01. URL: https://www.gnu.org/software/emacs/manual/elisp.html.

[11] M. Dennis Mickunas and John A. Modry. Automatic error recovery for LR parsers. *Communications of the ACM*, 21(6):459–465, 1978. doi:10.1145/359511.359519.

[12] Stefan Monnier and Michael Sperber. Evolution of Emacs Lisp. In *History of Programming Languages Conference*, 2020. Accepted.

[13] Masaru Tomita. *Efficient Parsing for Natural Language: A Fast Algorithm for Practical Systems*. Kluwer Academic Publishers, Norwell, MA, USA, 1985.

[14] Eric R. Van Wyk and August C. Schwerdfeger. Context-aware scanning for parsing extensible languages. In *Proceedings of the 6th International Conference on Generative Programming and Component Engineering*, GPCE '07, pages 63–72, October 2007. doi:10.1145/1289971.1289983.

[15] Tim A. Wagner and Susan L. Graham. Incremental analysis of real programming languages. In *Proceedings of the ACM SIGPLAN 1997 Conference on Programming Language Design and Implementation*, PLDI '97, page 31–43. ACM Press, June 1997. doi:10.1145/258916.258920.

[16] Tim A. Wagner and Susan L. Graham. Efficient and flexible incremental parsing. *Transactions on Programming Languages and Systems*, 20(5):980–1013, 1998. doi:10.1145/293677.293678.

[17] Mark N. Wegman. Parsing for structural editors. In *21st Annual Symposium on Foundations of Computer Science (sfcs 1980)*, pages 320–327, October 1980. doi:10.1109/SFCS.1980.33.

## About the author

**Stefan Monnier**  is a professor in the Département d'Informatique et Recherche Opérationelle of the Université de Montréal, where his research is focused on the design of programming languages and type systems. He was also the head maintainer of the Emacs text editor for several years.
Contact him at monnier@iro.umontreal.ca.